\documentclass[reprint,amsmath,amssymb,showpacs,aps,prl]{revtex4-2}
\usepackage{graphicx}
\usepackage[colorlinks,linkcolor=blue,citecolor=blue,urlcolor=blue]{hyperref}
\usepackage{multirow}
\usepackage{subfigure}

\makeatletter
\newcommand{\Rmnum}[1]{\expandafter\@slowromancap\romannumeral #1@}

\begin{document}

\title{Bloch-type photonic skyrmions in optical chiral multilayers}

\author{Qiang Zhang}
\author{Zhenwei Xie}
\email{ayst3\_1415926@sina.com}
\author{Luping Du}
\email{lpdu@szu.edu.cn}
\author{Peng Shi}
\author{Xiaocong Yuan}
\email{xcyuan@szu.edu.cn}

\affiliation{Nanophotonics Research Center, Shenzhen Key Laboratory of Micro-Scale Optical Information Technology $\&$ Institute of Microscale Optoelectronics, Shenzhen University, Shenzhen 518060, China}
\date{\today}

\begin{abstract}
Magnetic skyrmions are topological quasiparticles in magnetic field. Until recently, as one of their photonic counterparts, N\'eel-type photonic skyrmion is discovered in evanescent electromagnetic waves. The deep-subwavelength features of the photonic skyrmions suggest their potentials in optical imaging quantum technologies and data storage. So far, the Bloch-type photonic skyrmion has yet to be theoretically predicted and demonstrated. Here, by exploiting the photonic quantum spin Hall effect of a plasmonic vortex in tri-layered structure, we predict the existence of photonic twisted-N\'eel- and Bloch-type skyrmions in chiral materials.  Their chirality-dependent features can be considered as additional degrees-of-freedom for future chiral sensing, information processing and storage technologies.  In particular, our findings enrich the formations of photonic skyrmions and reveal a remarkable resemblance of the feature of chiral materials in two seemingly distant fields: photonic skyrmions and magnetic skyrmions. 
\end{abstract}

\maketitle

Skyrmions, `hedgehogs' of electron spins, are topologically stable magnetization swirls in magnetic materials. Typically, they are defined as two dimensional spin textures with non-zero topological winding numbers stablized at magnetic fields \cite{Bogdanov1989,Bogdanov1994}.  Their ultra-compact size and unique topological protections property make skyrmions promising for future applications in topological spintronics, information processing and data storage \cite{Parkin190,Kurumaji914,Foster2019,PhysRevX.10.031042,Yokouchi2020}, to name but a few. There are two typical magnetic skyrmions: N\'eel-type and Bloch-type skyrmions \cite{Muhlbauer2009,Yu2010,Heinze2011,Nagaosa2013,Kezsmarki2015,Wiesendanger2016,Fert2017,Bogdanov2020}. Until recently, the photonic counterpart of magnetic skyrmions has been discovered in surface plasmon polaritons (SPPs) \cite{Tsesses2018,Du2019}. To date, only N\'eel-type photonic skyrmion has been predicted and observed \cite{Tsesses2018,Du2019,Daviseaba2020,Tsesses2019,Bai:20,PhysRevLett.125.227201}. 

\begin{figure}[h!]
\centering
\includegraphics{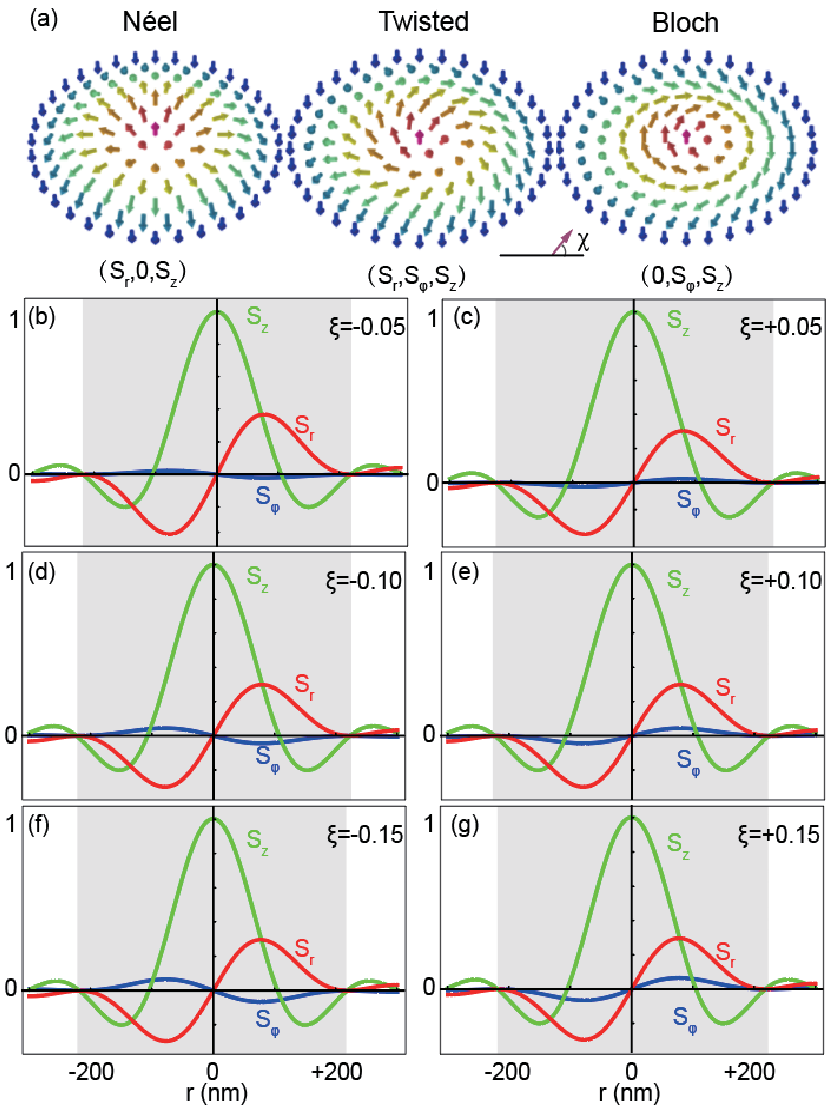}
\caption{(a) Photonic skyrmions can be defined according to their SAM verctor components, with $(S_r, 0, S_z)$ for N\'eel-type ($\vert \mathrm{cos}\chi\vert=1$), $(S_r, S_\varphi, S_z)$ twisted N\'eel-type  ($0<\vert \mathrm{cos}\chi\vert<1$) and $(0, S_\varphi, S_z)$ Bloch-type  ($\mathrm{cos}\chi=0$). So far only the N\'eel-type has been reported. (b-g) Evolution of the SAM vector components $S_r$, $S_\varphi$ and $S_z$ along the radius of the plasmonic-OV in a chiral-metal structure. Gray regions define the photonic skyrmion's size with $r=r_1$. }{\label{fig1}}
\end{figure}

In common optical materials (isotropic, nonmagnetic and nonchiral), SPPs solely exist for transverse magnetic (TM) polarization \cite{Maier2007}.  They are confined at the interface between a metal and a dielectic and exponentially decay in the perpendicular direction into the materials. SPPs carry transverse spin angular momentum (SAM) that is perpendicular to its momentum (propagation) direction due to the quantum spin Hall effect (QSHE) of light \cite{Bliokh1448}. For a plasmonic optical vortex (OV) beam at the metal - air interface, the transverse SAM ($S_r$) is perpendicular to the azimuthal direction of energy flux and thereby orients radially, as shown in Fig.~\ref{fig1}a. So its spin-orbit coupling results in a N\'eel-type skyrmion photonic structure with the local photon spin vector (that is, the orientation of the SAM) either along or opposite to the orbital angular momentum (OAM) \cite{Du2019}. The same mechanism applies to the photonic N\'eel-type skyrmion lattice generated by the interference of plane-wave SPPs \cite{Tsesses2018,Daviseaba2020}.  So far, both twisted N\'eel and Bloch-type skyrmions have been untapped in the novel domain of photonic skyrmions limited by this drawback.  Here we confine our discussion in the scheme of plasmonic OV and show how a Bloch-type photonic skyrmion can exist in chiral materials (materials that are noncentrosymmetric), and control its swirl handedness by tuning the chirality. 

In a plasmonic OV (Fig.~\ref{fig1}a), a N\'eel-type skyrmion can be understood according to its spin vector texture in the cylindrical coordinate ($S_r$, $S_\varphi$, $S_z$), with $S_\varphi=0$ and $\chi=0$, where $\chi$ is the \emph{helicity angle}.  By comparing with the Bloch-type  (0, $S_\varphi$, $S_z$), one finds immediately that the transverse spin $S_r$ should be zero. The wavevector of a surface wave (evanescent wave or SPPs) is complex and can be written in $\mathbf{k}=\mathbf{k_t}+i\mathbf{k_n}$ ($\mathbf{k_t}$ is tangential to the surface and related to the momentum, and $\mathbf{k_n}$ normal to the surface and related to the decay). The transversality condition $\mathbf{E}\cdot\mathbf{k}=0$ necessarily requires that the orthogonal components of the electric field vector in the propagation plane always have a $\pi/2$ phase difference and therby generates the transverse spin ($S_r$) due to the cycloidal rotation. It was shown that $S_r\propto\mathrm{Re}\mathbf{k}\times\mathrm{Im}\mathbf{k}$ \cite{Bliokh1448}, which indicates that the momentum (propogation) direction, the decay direction and the transverse spin direction of a SPP constitutes a right-handed system. This is the so-called \emph{spin-momentum locking} \cite{Bliokh1448,Bliokh,PhysRevLett.103.100401,PhysRevA.85.061801,Bliokh20151,VanMechelen:16}.    Notice that the SAM ($\bf{S} = \frac{\mathrm{1}}{\mathrm{4} \omega} \mathrm{Im}\left[ \epsilon_c \bf{E}^* \times \bf{E} + \mu_0 \bf{H}^* \times \bf{H} \right]$) continuously evolves in the material, so does the transverse spin $S_r$. So in principal zero-value point of $S_r$ exists in the case where $S_r$ varies from positive to negative. We construct this variation by exploiting the causality requirement and spin-momentum locking.  The causality requires that the decay of a surface wave must directs from the boundary toward the material. Hence, a confined struture (e.g. a slab) with its opposite surfaces bounded will introduce \emph{antiparallel} decay vectors $\mathbf{k_n}$, while the momentum vector $\mathbf{k_t}$ inside is constant. As such, according to the right-handed spin momentum locking, $S_r$ flips sign at the two boundaries. Consequently, a zero $S_r$ will emerge between the boundaries. 
Besides, study highlights that in chiral materials, SPPs exhibit hybrid modes of TM and transverse electric (TE) polarizations \cite{Engheta:89}. This offers the opportunity to form a nonzero $S_\varphi$. Therefore we show Bloch-typed photonic skyrmion can be realized in a tri-layered metal-insulator-metal (MIM) structure with the introduction of chirality. 

Electromagnetic response of chiral material obeys the following constitutive relation \cite{LSTV1994}: ${\mathbf D}=\epsilon_c{\mathbf E}+\mathrm{i}\xi\sqrt{\mu_0\epsilon_0}{\mathbf H}$, ${\bf B}=\mu_0{\mathbf H}-\mathrm{i}\xi\sqrt{\mu_0\epsilon_0}{\mathbf E}$, where $\xi$ and $\epsilon_c$ are the chirality parameter and the permittivity of the chiral material, respectively.  $k_\pm=(\sqrt{{\epsilon_c}/{\epsilon_0}}\pm\xi)\omega/c$ are eigen-wavenumbers corresponding to the right-handed circularly polarized (RCP) and left-handed circularly polarized (LCP) waves, respectively. Using Bohren decompositions \cite{Bohren1974}, $\mathbf{F}_{\pm}=\mathbf{E}{\pm}\mathrm{i}\eta\mathbf{H}$ ($\eta=\sqrt{\mu_0/\epsilon_c\epsilon_0}$ is the wave impedance of the chiral medium), the field equations can be written as $\nabla\times\mathbf{F}_{\pm}={\pm}k_{\pm}\mathbf{F}_{\pm}$, $ 
\nabla^2\mathbf{F}_{\pm}+k_\pm^2\mathbf{F}_{\pm}=0$, with $\mathbf{E}=(\mathbf{F}_++\mathbf{F}_-)/2$ and $\mathbf{H}=(\mathbf{F}_+-\mathbf{F}_-)/2\mathrm{i}\eta$. 
${\mathbf F}_{\pm}$ field of a vortex beam in the chiral medium has mode solution as
${{\mathbf{F}}_ \pm } = {{\mathbf{f}}_ \pm }\left( r \right)\exp \left[ {\mathrm{i}\left( {\beta z - \omega t + l\varphi } \right)} \right]$, where $l$ represents the topological charge. In the cylindrical coordinate system, $f_{\pm z}$ is determined by $\frac{1}{r}\frac{\partial }{{\partial r}}\left( {r\frac{{\partial {f_{ \pm z}}}}{{\partial r}}} \right) + \left( {k_ \pm ^2 - \beta ^2 - \frac{{{l^2}}}{{{r^2}}}} \right){f_{ \pm z}} = 0$, which has the solution $f_{\pm z} = a_{\pm}J_{l}\left(k_r r\right)$. Here  $k_r ^2=k_{\pm}^{2}-\beta^{2}$, $a_{\pm}$ are arbitrary constants representing the field amplitude, and $J_{l}$ is the $l$-th order Bessel function of the first kind. The rest of the field functions are given by the curl of $\mathbf{F_\pm}$.
For a plasmonic OV, the in-plane wavevector component ($k_r$) is larger than the wavevector ($k_{\pm}$) of the beam, resulting in an imaginary axial wavevector component (denoted as $\mathrm{i}k_{+z}$ and $\mathrm{i}k_{-z}$, representing RCP and LCP eigenwaves respectively). By replacing the $z$-component wavevector in the equations above with the imaginary value $\mathrm{i}k_{+z}$ and $\mathrm{i}k_{-z}$, we can obtain the $f$-fields for a plasmonic OV in the chiral medium. 
We can treat the metal in the same way. After obtaining the electromagnetic field solutions in the metal, we get the dispersion relation of the plasmonic OV from the eigen-solution for the boundary condition at the chiral - metal interface (see the Supplemental Material for details). Subsequently, $k_r$ is determined by the dispersion relation and the eigen-field solutions can be resolved. Consequently, the SAM of the plasmonic OV can be calculated by $\bf{S} = \frac{\mathrm{1}}{\mathrm{4} \omega} \mathrm{Im}\left[ \epsilon_c \bf{E}^* \times \bf{E} + \mu_0 \bf{H}^* \times \bf{H} \right]$, and is written as follows: 
\begin{equation}
\mathbf{S}=\begin{pmatrix} S_r \\ S_\varphi \\ S_z \end{pmatrix} =\frac{\epsilon_c}{\mathrm{4} \omega} \frac{l}{r} \begin{pmatrix}  \frac{J_{l}^2}{k_r^2} \left( {a_+^2} k_{+z} + {a_-^2} k_{-z} \right)   \\ \frac{J_{l}^2}{k_r^2} \left({a_+^2} k_{+} - {a_-^2} k_{-} \right) \\ \frac{J_{l} J_{l}^\prime}{k_r} \left({a_+^2}  + {a_-^2}  \right)\end{pmatrix} {\label{eq1}}
\end{equation}

\begin{figure}[htbp!]
\centering
\includegraphics{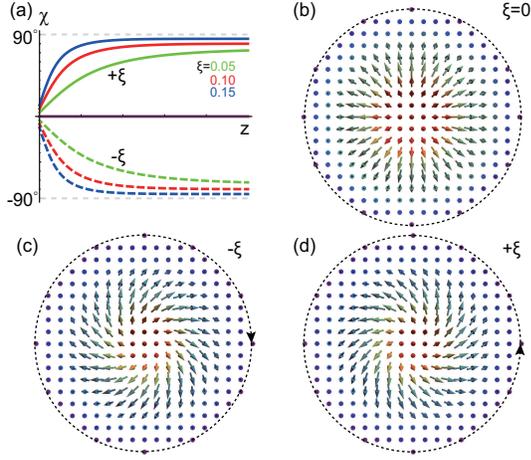}
\caption{(a) Helicity angle ($\chi$) profile as a function of $z$ for the twisted skyrmions with different chirality parameters. $\chi$ is unable to reach $90^\circ$ inside the chiral material regardless of its chirality due to the existence of nonzero $S_r$. (b-d) Illustrations of the skyrmion structures at the interface. The handedness of the twist depends solely on the sign of the material's chirality.}{\label{fig2}}
\end{figure}
In the isotropically nonchiral system, SPPs solely exist for TM polarizations, and as a result $S_\varphi$ is constantly zero. Compared to the spin texture expression for air - metal structure \cite{Du2019}, we can see that a nonzero $S_\varphi$ is emerged in Eq.~\eqref{eq1}. Most importantly this term is chirality dependent. According to the reciprocity, a LCP wave in a right-handed system undergoes same physical response with a RCP wave in a left-handed system. $S_\varphi$ is therefore an odd function since $a_+^2k_+$ and $a_-^2k_-$ would swap places for opposite chirality in this expression, which is also confirmed by our numerical calculations. We hence plot the spin vector structure's profile of the plasmonic OV in Fig.~\ref{fig1}. Parameters throughout the paper are chosen as follows: $l=1$, $\lambda=532$ nm, $\epsilon_c=1.33^2$ and Drude permittivity model for gold. For $\xi=0$, the spin texture for the SAM is a photonic analogue of a N\'eel-type magnetic skyrmion \cite{Kezsmarki2015}, as shown in Fig.~\ref{fig2}b. For nonzero $\xi$, the spin texture is a photonic analogue of a twisted N\'eel-type magnetic skyrmion \cite{PhysRevB.87.094424,PhysRevB.90.014406,PhysRevLett.117.087202,PhysRevLett.120.227202}, as shown in Fig.~\ref{fig1}b-g and Fig.~\ref{fig2}c-d, which has never been reported yet.

\begin{figure}[htbp!]
\centering
\includegraphics{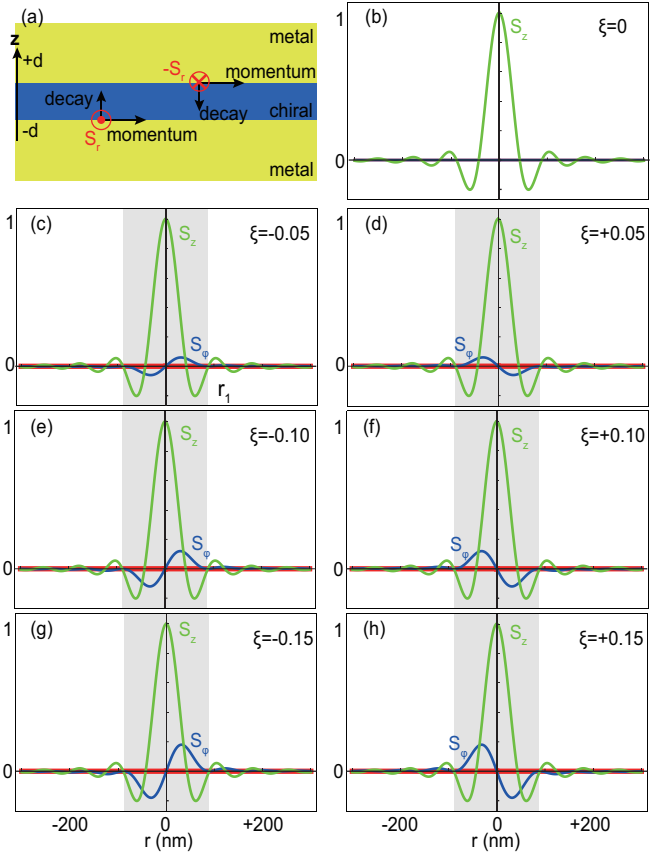}
\caption{(a) Schematic illustration of the opposite transverse spins ($S_r$) on the surfaces of a film that supports coupled plasmons. (b) For a nonchiral system ($\xi=0$), the SAM in the center of the film only has $S_z$ component, so it is not a skyrmion. (c-h) $S_\varphi$ emerges when the chirality is introduced and depends on the magnitude and sign of $\xi$. }{\label{fig3}}
\end{figure}
To further verify the skyrmion analogy, we calculate the skyrmion number (n) of the spin vector structure $n=\frac{1}{4\pi}\int\int\mathbf{M}\cdot\left(\frac{\partial\mathbf{M}}{\partial x}\times\frac{\partial\mathbf{M}}{\partial y}\right)dxdy$, where $\mathbf{M}=\left(\cos\phi(\alpha) \sin\theta(r),\sin\phi(\alpha) \sin\theta(r),\cos\theta(r)\right)$ represents the unit vector in the direction of the local SAM vector within the plasmonic OV ( $\mathbf{r}=\left(r\cos\alpha,r\sin\alpha\right)$), and the integral is taken over a unit cell of the skyrmion in the $X$-$Y$ plane. Thus the skyrmion number results in $n=-\frac{1}{4\pi}\ [\cos\theta\left(r\right)]_{\theta\left(0\right)}^{\theta\left(r_1\right)}[\phi\left(\alpha\right)]_{\left(\alpha=0\right)}^{\left(\alpha=2\pi\right)}$. For a twisted spin texture, $\phi\left(\alpha\right)=\alpha+\chi$ \cite{Nagaosa2013}, where $\chi$ is the helicity as defined in Fig.~\ref{fig1}a.  Eq.~\ref{eq1} implies that the helicity is independent of the radius as $\chi=S_\varphi/S_r$. Therefore, for a photonic spin structure with the spin vector changing progressively from the `up' state ($\cos\theta\left(0\right)=1$) to the `down' state ($\cos\theta\left(r_1\right)=-1$) at each domain as shown in the gray zone of Fig.~\ref{fig1}b-g, the skyrmion number can be evaluated as $n = 1$.

At the chiral-metal interface, the plasmonic OV possesses twisted skyrmion spin texture in analogy to the \emph{surface twisted skyrmions} in bulk chiral magnets \cite{PhysRevB.87.094424,PhysRevB.90.014406,PhysRevLett.117.087202}. Inside the chiral magnets, Bloch-type magnetic skyrmions with $\chi=\pm90^{\circ}$ are stable and uniformely stacked along the bulk crystal \cite{PhysRevLett.117.087202,PhysRevLett.120.227202}.
However, it should be stressed that for a chiral plasmonic OV at single interface, Bloch-type skyrmions do not exist whatsoever in the chiral material. Because regardless of the distance from the chiral surface, a plasmonic OV always carrys a nonzero $S_r$ component even when the chirality is strong. We illustrate the quantitative evolution of the helicity angle $\chi$ along the z direction of a plasmonic OV in such a chiral-metal structure, in Fig.~\ref{fig2}a. $\chi$ exponentially approaches to a stable angle as the plasmonic OV decays into the chiral material and eventually cutoff before $\pm90^{\circ}$. The cutoff angle depends on the chirality. However, the mechanism for the formation of such a twisted photonic skyrmion is prerequisite to the realization of a Bloch-type photonic skyrmion.
\begin{figure}[htp!]
\centering
\includegraphics{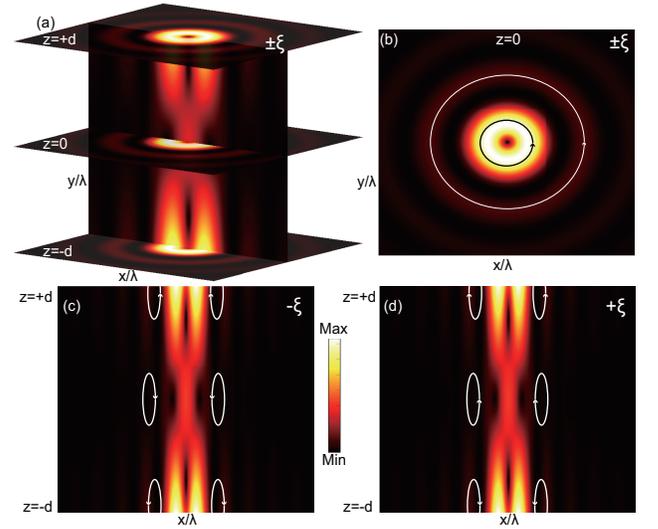}
\caption{(a) Energy flux inside the chiral film ($-d\leq z \leq +d$). Nonzero $P_z$ exists. (b) Azimuthal energy flux is dominant and is independent of the material's chirality. (c, d) $P_z$ and $P_r$ have reversed signs for $\xi$ with opposite signs, implying the chirality-dependent behavior of $S_\varphi$ via $\mathbf{S}\propto\nabla\times\mathbf{P}$.}{\label{fig4}}
\end{figure}

We have proved that in a single interface structure, the photonic Bloch-type skyrmion does not exist even for systems with strong chirality. Specifically, this is restricted by the intrinsic QSHE of the SPPs \cite{Bliokh1448}, that is the transverse spin ($S_r$) never vanishes in such a structure. Such effect hinders the formation of Bloch-type photonic skyrmions. Nevertheless, we can in turn exploit this effect by engineering \emph{counter transverse spin} structures to extinguish $S_r$ while keeping $S_\varphi$ component to realize $\chi=\pm90^\circ$. So in a MIM structure (Fig.~\ref{fig3}a), the SPPs have transverse spins with opposite signs on the surfaces of the core film. The SAM expression for the plasmonic OV in such a structure ($-d\leq z \leq +d$) is (see the Supplemental Material for details):
\begin{figure*}[htbp!]
\centering
\includegraphics{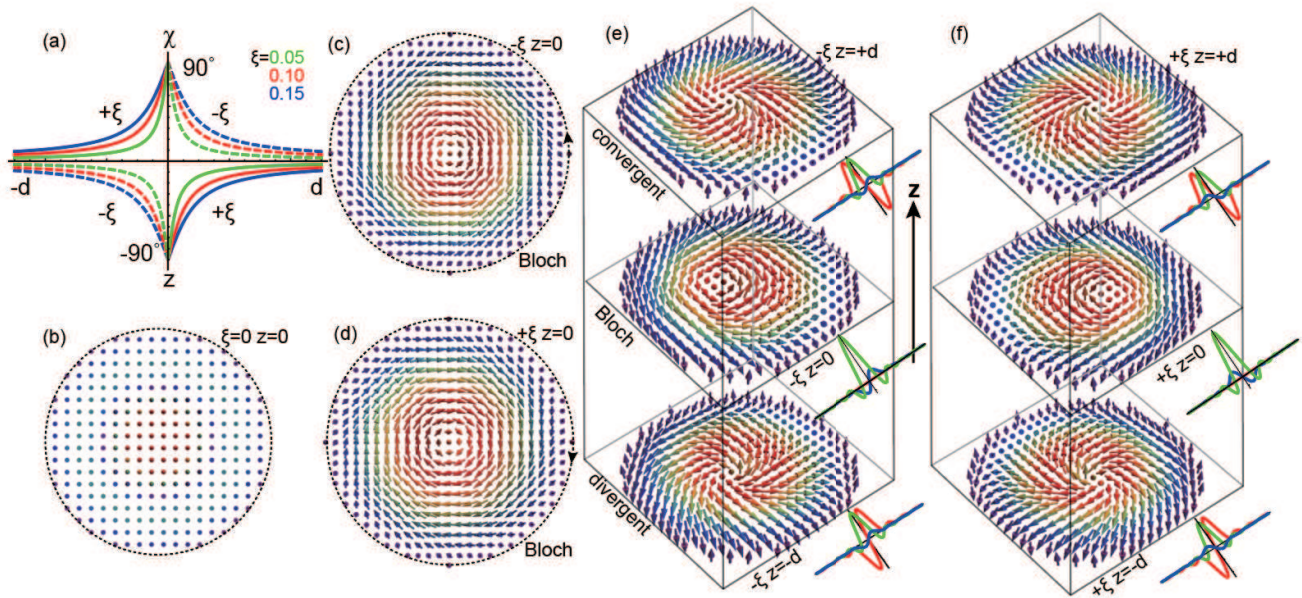}
\caption{Illustrations of representative spin textures.  (a) Diagram for the evolution of the helicity angle of the skyrmion's spin texture as a function of the chiral film thickness. (b) SAM vectors with only $S_z$ component in the center of a nonchiral film. (c, d) Bloch-type skyrmion structures with opposite handedness of swirls in the chiral films with opposite chirality, respectively. (e,f) Twisted skyrmion structures on the bottom and top surfaces of the chiral film, respectively. The insets show the corresponding ($S_r$, $S_\varphi$, $S_z$) distributions.}{\label{fig5}}
\end{figure*}
\begin{equation}
\mathbf{S}
= \frac{\epsilon_c}{\mathrm{4} \omega} \frac{l}{r}\begin{pmatrix} 
\frac{J_{l}^2}{k_r^2} [( -{a_{1+}^2}e^{2k_{+z}(z-d)} + {a_{2+}^2}e^{-2k_{+z}(z+d)})k_{+z}
\\ - ({a_{1-}^2}e^{2k_{-z}(z-d)} + {a_{2-}^2}e^{-2k_{-z}(z+d)})k_{-z}]
\\
  \\  \frac{J_{l}^2}{k_r^2} [(a_{1+}e^{k_{+z}(z-d)} + a_{2+}e^{-k_{+z}(z+d)})^2k_{+}
\\- (a_{1-}e^{k_{-z}(z-d)} + a_{2-}e^{-k_{-z}(z+d)})^2k_{-}] \\
\\
 \frac{J_{l} J_{l}^\prime}{k_r^3}[(a_{1+}e^{k_{+z}(z-d)} + a_{2+}e^{-k_{+z}(z+d)})^2k_+^2
\\+ (a_{1-}e^{k_{-z}(z-d)} + a_{2-}e^{-k_{-z}(z+d)})^2k_-^2
\\+(a_{1+}e^{k_{+z}(z-d)} - a_{2+}e^{-k_{+z}(z+d)})^2k_{+z}^2
\\+ (a_{1-}e^{k_{-z}(z-d)} - a_{2-}e^{-k_{-z}(z+d)})^2k_{-z}^2] \end{pmatrix} {\label{eq2}}
\end{equation}

When the SPPs decay inside the core, a region of $S_r=0$ will be emerged somewhere (this region is exactly the center of a symmetric structure for instance as in the following discussion, asymmetric structure example can be found in the Supplemental Material). Here we take a core of thickness $2d=10$ nm as an example, while the conclusion holds consistent for other thickness (see Sec.\Rmnum{4} in the Supplemental Material). For a nonchiral and isotropic material, this results in a plasmonic OV with spin texure which has solely $S_z$ component in the structure center, as shown in Fig.~\ref{fig3}b and Fig.~\ref{fig5}b. When chirality is introduced, a nonzero $S_\varphi$ component appears whose magnitude/sign are associated with the magnitude/sign of the material's chirality, see Fig.~\ref{fig3}c-h. It is noticed that compared to the single interface case (Fig.~\ref{fig1}h for instance), the handedness of the Bloch-type skyrmion's swirl (Fig.~\ref{fig3}h for instance) is reversed for the same material chirality. This results from the strong field coupling between the top and bottom surfaces of the chiral material when it is a very thin film such as 10 nm. The strong coupling magnifies the discrepancy between coefficients $a_+$ and $a_-$ associated with the RCP and LCP eigen waves and thereby reverses the sign of the term ``$a_+^2k_+-a_-^2k_-$''. When the centeral film thickness is increased, the coupling attenuates and $S_\varphi$ behaves like the single interface structure while $S_r$ maintains zero in the center (see Sec.\Rmnum{5} in the Supplemental Material). It is also demonstrated that a more deeply subwavelength skyrmion size ($r\sim$ 90 nm, 1/6 $\lambda$) is achieved compared with that ($r\sim$ 220 nm) in the single interface structure.  The compressed effective wavelength by such a highly confined structure contributes to a ultracompact plasmonic OV together with the skyrmion structure.  

On the other hand, the Poynting vector (that is, the energy flux density) of the plasmonic-OV can be calculated by $\mathbf{P} = \mathrm{Re}(\mathbf{E}^* \times \mathbf{H})/2$ (see the Supplemental Material for details) and has shown its relation to the SAM via $\mathbf{S}\propto\nabla\times\mathbf{P}$ \cite{arXivShi,Shi2020}. A more close inspection at the energy flux corresponding to the Bloch skyrmions in Fig.~\ref{fig3}g and Fig.~\ref{fig3}h is plotted in Fig.~\ref{fig4}. Inside the chiral film, a nonzero $P_z$ component exists (Fig.~\ref{fig4}a, c and d) which is distinctively different from a nonzero plasmonic-OV as in ref.~\cite{Du2019}. Azimuthal energy flux $P_\varphi$ still dominates (Fig.~\ref{fig4}b). The reversed $P_z$ component (Fig.~\ref{fig4}c and Fig.~\ref{fig4}d) also implies that the azimuth component ($S_\varphi$) of the SAM exhibits opposite signs for $\pm\xi$.

Finally, we reveal an intruiging resemblance of the characteristics of the photonic skyrmions and magnetic skyrmions in chiral materials. The evolution of the helicity angle of a skyrmion's spin texture along the chiral film thickness is presented in Fig.~\ref{fig5}a. All the representative spin textures can be described from this diagram. For a nonchiral film, the spin vectors in the center are depicted in Fig.~\ref{fig5}b. The photonic skyrmions in a plasmonic chiral multilayerd structure (three-layered metal-chiral-metal structure in this work) possess numerous similarities with the magnetic skyrmions in chiral magnets \cite{PhysRevLett.117.087202,PhysRevLett.120.227202} as follows: on the top and bottom surfaces of a chiral film, divergent and convergent (Fig.~\ref{fig5}e and Fig.~\ref{fig5}f) twisted N\'eel-type skyrmions are formed, respectively; and eventually the twisted skyrmions evolves into a Bloch-type skyrmion in the chiral film center, with its swirl handedness solely depends on the film's chirality (Fig.~\ref{fig5}c and Fig.~\ref{fig5}d).  

In summary, we have demonstrated theoretically the existence of a Bloch-type photonic skyrmion. Without loss of generality, we first take a chiral-metal plasmonic interface as an example and shows how a twisted N\'eel-type photonic skyrmion structure which had never been reported is formed in a plasmonic OV. The twist of skyrmion's swirl handedness is determined by the magnitude and sign of the material's chirality. Based on this finding, we subsequently present the mechnism of extinguishing the transverse SAM ($S_r$) of the plasmonic OV inside a three-layered metal-chiral-metal structure. The Bloch-type photonic skyrmion is predicted in this structure and the skyrmion size is greatly minimized  compared to the single-interface structures. The swirl handedness of Bloch-type photonic skyrmions is determined by the magnitude and sign of the material's chirality simultaneously. This property implies a possible chiral sensing scheme in optics. We note that our theory can be extended to multilayered thin-film plasmonic structures with successive interfaces that will further enhance the $S_\varphi$ component, similar to the enhanced coupling of magnetic skyrmions in the successive ultrathin Co layers \cite{Luchaire2016,Soumyanarayanan2017}. We also note that it is possible to find Bloch-type electric skyrmion (like the N\'eel-type electric skyrmion in ref.\cite{Tsesses2018}) in structured chiral multilayers relying on the hybrid mode property of chiral SPPs. Therefore, our study enlarges the family of photonic skyrmions. Additionally, our findings may pave the way to simple-shaped nanophotonic skyrmionic devices and offer more degrees-of-freedom for future sensing, information processing and storage technologies.
\\
\\

\begin{acknowledgments}
This research was supported by the National Key R$\&$D Program of China, 2018YFB1801801, National Natural Science Foundation of China (61935013, U1701661, 61975133, 11774240, 61622504, 11604218 and 61705135), leadership of Guangdong province program grant 00201505, the Natural Science Foundation of Guangdong Province grants 2016A030312010, 2020A1515011185, the Science and Technology Innovation Commission of Shenzhen grants KQTD2017033011044403, KQJSCX20170727100838364, ZDSYS201703031605029, JCYJ20180507182035270 and JCYJ20200109114018750, Shenzhen University (2019075). QZ also acknowledges the funding support by the NSFC 12047540.

\end{acknowledgments}

\bibliographystyle{apsrev4-2}
\bibliography{apssamp}
\end{document}